\documentclass[12pt]{article}

\usepackage{lipsum} 
\usepackage{graphicx} 
\usepackage{amsmath} 
\usepackage{amsfonts} 
\usepackage{amssymb} 
\usepackage{hyperref} 
\usepackage{cleveref} 
\usepackage{xcolor}
\usepackage{soul}
\usepackage{authblk}

\title{Axial Confinement in the Novatron \\ Mirror Machine}

\author[1,2]{J. Scheffel}
\author[1]{J. J\"aderberg}
\author[1]{K. Bendtz}
\author[1]{\authorcr R. Holmberg}
\author[1]{K. Lindvall}

\affil[1]{\small{Novatron Fusion Group, Teknikringen 31, Stockholm, Sweden}}
\affil[2]{\small{Electromagnetic Engineering and Fusion Science, KTH Royal Institute of Technology, Teknikringen 31, Stockholm, Sweden}}

\date{} 

\begin{document}

\maketitle

\begin{abstract}
\noindent The Novatron magnetic mirror fusion reactor concept features significant advantages. These include stability against MHD interchange and kinetic DCLC modes, axisymmetry, and minimized radial particle drifts and neoclassical losses. For achieving a ratio $Q \ge 30$ of fusion power to heating power, axial particle confinement is uniquely designed to rely on the simultaneous use of three distinct forces; magnetic mirrors, electrostatic potentials, and ponderomotive forces in a tandem-like configuration. Axial confinement physics theory is here analyzed and compared to earlier mirror configurations. Scenarios for D-T, D-D, and catalyzed D-D fusion plasmas are outlined.   \\
\end{abstract} 

\section{Introduction}
\noindent Axial confinement of particles in open-ended systems for fusion has been a pivotal topic since the development of magnetic mirror machines in the early 1950s \cite{Fowler:1981,Baldwin:1977}. It has a profound impact on the machine's accessible Q-value; the ratio between produced fusion power and plasma heating power. 

The mirror fusion program in the USA was discontinued in the mid-1980s, largely due to the fact that all the proposed solutions to the axial confinement problem instead resulted in significant plasma losses or instabilities. There  was no shortage of ideas. However, it was soon recognized (see Section 2), that the Q-value for a simple mirror, where plasma is confined in an axial magnetic field with magnetic mirrors placed at each end, could not extend much beyond 1. It proved impossible to attain values around 30, which are required for a power plant. 

As a consequence, the concept of the tandem mirror was suggested in 1976 \cite{Fowler:1977, Dimov:1976}. In this configuration, the central plasma cell was complemented by one additional magnetic mirror plug cell at each end of the central cell. By enhancing the positive electrostatic potential of the plug cells, central cell ions would be electrostatically confined axially, allowing for a higher Q-value. The Boltzmann relation for electrons shows that there are two ways to attain this potential; either by heating the plug cell electrons to a temperature above that of the central cell electrons, or by significantly increasing (up to a factor 10) the density in the plug cell in relation to that of the central cell. The colder central cell electrons were, however, expected to interact with the plug cell electrons. Consequently, the latter approach was chosen. It quickly became evident that the associated high densities require very strong magnets as well as very high energy (on the order of MeV) neutral beam heating in a full-scale power plant. A possible solution was presented in 1979 \cite{Baldwin:1979} whereby a ``thermal barrier'' thermally isolated the plug cell from the central cell electrons. The plug cell electrons could then be heated to the required level in order to maintain a high positive electrostatic potential in the plug cells. This approach, however, turned out to be successful only for low (order $10^{18}$ $\text{m}^{-3}$) central cell densities, for which the trapping of passing ions did not overwhelm the barrier \cite{Post:1987,Simonen:1985}.

The problem of axial particle confinement has been addressed also by the use of so-called ponderomotive forces (``RF plugging''), which act repulsively on both electrons and ions \cite{Sato:1982, Okamura:1984,Sato:1985,Fujita:1988}. In the experiments at RFC-XX in Nagoya, Japan, an axially symmetric plasma was confined by magnetic cusps at both ends. The unstable central cell plasma was stabilized by the favourable field line curvature in the cusp fields. A ten-fold increase in confinement time when using ponderomotive plugging was reported \cite{Sato:1986}.

In this article we investigate axial particle confinement in the Novatron magnetic mirror concept \cite{Conceptpaper2024} in order to estimate obtainable Q-values. The Novatron magnetic field is axially symmetric, thus all particle drifts are azimuthal. Radial, neoclassical losses are expected to be small. A fundamental feature of the Novatron is favourable magnetic curvature in the plasma region, stabilizing MHD interchange modes, and eliminating the need for magnetic fields outside the central plasma to be tailored for stability. The design also mitigates so-called drift cyclotron loss cone (DCLC) micro-instabilities, traditionally constituting a substantial problem for mirror machines. Detailed studies of how to address Novatron macro- and microinstabilities are under way \cite{Conceptpaper2024} and will be presented in separate articles. 

The first experimental platform featuring a Novatron magnetic configuration, N1, is currently being completed and prepared for experiments at the Alfv\'en Laboratory at KTH, Stockholm, Sweden. The N1 platform is a 3.8 m tall, 1.8 m wide central cell including expanders. The plasma radius is 0.33 m, and the mirror-to-mirror distance is 1.3 m. The main purpose of the N1 is to experimentally validate the stability of the central cell against interchange modes, as well as act as a test bed for RF ponderomotive plasma plugging. A second experimental platform, the N2, is currently in the design phase.  Approximately twice as large as the N1 and equipped with tandem mirrors, it is aimed at testing and developing Q-enhancement techniques.

\section{Q-value of simple mirrors}
\noindent Of major interest for all fusion confinement schemes are the attainable Q-values. Scientific Q-value refers to the ratio between generated fusion power and the input power used to heat the plasma. For a power plant, where the generated electricity should also power associated electrical and mechanical systems, Q is required to be about 30 or more. In this section, we will discuss attainable Q-values in simple, traditional mirror machines as well as for a basic Novatron geometry. Finding that these Q-values are insufficient, the next section is devoted to Q enhancement measures, after which tandem mirror geometry will be discussed. As will be seen, in tandem geometry the simultaneous use of magnetic mirrors, electrostatic plasma potentials and ponderomotive forces, the central cell ion confinement time, and thus also that of the electrons, can be significantly enhanced.

A detailed study of plasma confinement in an open-ended mirror machine requires that collisions, which lead to loss-cone degradation, need to be taken into account, using a Fokker-Planck analysis. This was done by Pastukhov \cite{Pastukhov:1974} and extended by others \cite{Cohen:1978, Rognlien:1980, Najmabadi:1984}, who derived explicit, approximate formulas for the electron and ion confinement time when there is a confining, electrostatic plasma potential present. Following Pasthukov's approach, we will in this study base the axial confinement analysis on the parallel velocities of ions and electrons. Our motivation is that, for particle confinement in a collisionless fusion plasma, the relation $\tau_{\alpha \alpha} \gg \tau_{L\alpha} $ must hold, where $\tau_{\alpha \alpha}$ is the self-collision time for species $\alpha$ (ions or electrons), and $\tau_{L\alpha}$ is the time for the confined particle to traverse between the two mirror regions at distance $L$ apart. The particles of a fusion plasma bounce many times between the mirrors before colliding, and we can therefore safely apply this assumption. As a consequence, we can use the fact that both particle energy and adiabatic moment $\mu$ are conserved during the motion for confined particles, leading to simple, explicit formulas for $v_{\alpha \parallel}$. For confinement, it must hold that $v_{\alpha \parallel}$ changes sign somewhere in the plasma domain. This can be due to magnetic mirroring, electrostatic confinement, ponderomotive RF plugging, or a combination thereof.

To confirm that $\tau_{\alpha \alpha} \gg \tau_{L\alpha} $ holds, we write $\tau_{L\alpha} = L/v_{T \alpha}$, where $v_{T \alpha}=(2eT_\alpha/m_\alpha)^{1/2}$ is the thermal velocity of species $\alpha$ with temperature $T_\alpha$ in eV. Thus we write (see \cite{NRLPlasmaFormulary})
\begin{equation}
	\tau_{ee} / \tau_{Le} = 3.3 \cdot 10^{11} \frac {T_e^{3/2}} {n \ln \Lambda} / \tau_{Le} = 2.0 \cdot 10^{17} \frac {T_e^2}{L n\ln  \Lambda}
\end{equation}
\begin{equation}
	\tau_{ii} / \tau_{Li} = 2.1 \cdot 10^{13} \frac {T_i^{3/2}} {n \ln  \Lambda} \sqrt \frac {m_{i}}{m_p} / \tau_{Li} = 2.9 \cdot 10^{17} \sqrt{\frac{m_i}{m_p}}\frac {T_i^2}{L n\ln  \Lambda}
\end{equation}

\noindent for electrons and singly charged ions with mass to proton mass ratio $m_i/m_p$, respectively. Here $n$ denotes number density and $\ln  \Lambda$ the Coulomb logarithm. For a hydrogen plasma with $T_e = T_i = 20$ keV, $\ln  \Lambda =19$ and $n = 1 \cdot 10^{20}$ m$^{-3}$  in a Novatron with $L=5$ m, we obtain $\tau_{ee} / \tau_L = 8.2 \cdot 10^3$ and $\tau_{ii} / \tau_L = 1.2 \cdot 10^4$, validating our assumption. It may be noted, however, that for a laboratory plasma with, for example, $T_e = T_i = 20$ eV, these ratios are much smaller than unity, implying that the particles neither conserve energy (due to collisions) nor adiabatic moment $\mu$, excluding $v_\parallel$  as a tool for estimating confinement.

\subsection{Simple classical mirror, initial state}
The ``simple'' mirror machine, being investigated since the early 1950s, consists of a plasma confined in an axial magnetic field between two magnetic mirrors. Plasma confinement in this configuration may be stationary, but not static, due to continuous plasma leakage through the mirrors. The energy confinement time in this quasi-steady state will be insufficient for a fusion power plant. This argument will now be physically motivated.

During plasma start-up, electrons and ions will, in the absence of collisions, move along the magnetic field $\textbf B$, pointing in the $z$-direction at the midplane, according to (see Appendix A)
\begin{equation}\label{simple-mirror-parallel-velocity}
	v_{\alpha \parallel} = v_{\alpha 0} \sqrt {1-\frac{B}{B_0} \sin^2 \nu} \quad .
\end{equation}

\noindent Here subscript ``$\alpha$'' denotes particle species (electrons or ions) and ``$0$'' refers to values at the $z=0$ midplane. Thus $m_{\alpha}v_{\alpha 0}^2/2$ is the particle kinetic energy, $B_0$ is the magnetic field amplitude, and $\nu$ is the pitch angle at $z=0$. Constancy of the adiabatic moment $\mu$ is assumed when deriving Eq. (3).

All particles, electrons and ions alike, for which
\begin{equation}
	B=B_M > B_0 / \sin^2 \nu 
\end{equation}

\noindent will bounce at the magnetic mirrors where the maximum field is $B_M$. Other particles will be lost, as indicated by the so-called loss-cone angle given by
\begin{equation}
	\nu < \arcsin (B_0/B_M) \equiv \arcsin(1/R_M) .
\end{equation}

\noindent where $R_M \equiv B_M/B_0$ is the mirror ratio.

\subsection{Simple mirror, ambipolar state}
Electrons move faster than the ions and are initially lost more rapidly from the central mirror. Their shorter collision time $\tau_{ee} \ll \tau_{ii}$ causes them to be collisionally spread into the loss cone on a shorter time scale than the ions. As a result, a positive, ambipolar electrostatic potential $\phi$ evolves in the plasma between the $z=0$ plane and the mirrors. This potential will increase with time until it is large enough to pull back most electrons in order to preserve quasi-neutrality on a time scale related to the slow ion loss through the mirrors. We first consider the parallel particle motion in presence of the ambipolar potential and, in the following section, we proceed to discuss axial electron and ion confinement times, based on analytical solutions of the Fokker-Planck equation.

\subsubsection {Parallel particle motion}
 Following the arguments outlined in the preceding paragraph, we may expect that at the mirrors, with a constant $C_{\phi} \approx 3-6$ (computed in the next Section), it holds that
\begin{equation}
	e(\phi_0 - \phi) \equiv e\Delta \phi = e(\phi_0 - \phi_M) = C_{\phi} T_e \hspace{0.25em} ,
\end{equation}
\noindent where \( T_e \) is in eV with the factor \( e \) omitted here and below. This relation holds independently of $R_M$, since the potential drop $\phi_0-\phi_M$ must stop essentially all electrons from escaping faster than the ions. 

The $z$-dependence of the electrostatic potential is related to the plasma density $n$ through the Boltzmann equation for electrons;
\begin{equation}
	n = n_0 e^{-e(\phi_0-\phi)/{T_e}} = n_0 e^{-e\Delta \phi/{T_e}} \hspace{0.25em} .
\end{equation}
\noindent Note that the density profile, and thus $\phi$, is essentially constant along $z$, only to drop to low values in the vicinity of the magnetic mirror, where the density is low; $n_M \ll n_0$. The low density near the mirrors has its cause from two effects. From a particle perspective, conservation of the magnetic moment $\mu$ increasingly prevents particles from reaching the strong-field mirror region. From a plasma perspective, on the other hand, equilibrium requires that the sum of plasma kinetic and magnetic pressures is conserved, with the resulting effect of decreased parallel plasma pressure towards the magnetic mirrors. Thus it holds that $\phi_0 \gg \phi_M$, and that $\Delta \phi > 0$ away from the midplane $z=0$. Since $\phi$ is everywhere positive, with a gradient near the mirrors, it must hold that
\begin{equation}
	T_e \ll e\Delta \phi < T_i
\end{equation}

\noindent in order to avoid significant electrostatic expulsion of ions; if $T_e \approx T_i$, it would also hold that $e\Delta \phi \gg T_i$ which is incompatible with ion confinement. Due to that $\tau_{ee} \ll \tau_{ii}$, electrons are assumed Maxwellian. But if the ions cannot be assumed to be Maxwellian due to, for example, neutral beam heating, $T_i$ should be replaced by a characteristic ion energy $E_i$  \cite{Fowler:1981}. The inequality (8) explains why $T_e \ll T_i$ must be upheld in the classic mirror, which unfortunately conflicts with the requirement of long pulses and, hence, temperature equilibration in a fusion-relevant plasma.

Including the ambipolar potential and assuming a steady state, Eq. (3) generalizes to (see Appendix A)
\begin{equation}
	v_{\alpha \parallel} = v_{\alpha 0} \sqrt {1-\frac{B}{B_0} \sin^2 \nu + q_\alpha \Delta \phi /(m_{\alpha}v_{\alpha 0}^2/2)}.
\end{equation}

\noindent Here $q_\alpha =(-e,e)$ for (electrons, ions), respectively. Since $\Delta \phi > 0$ in the simple, ambipolar mirror, the electrostatic term \textit{decreases $v_{\parallel}$ for electrons and increases $v_{\parallel}$ for ions}. It is seen that whereas all but a small population of highly energetic electrons find it hard to escape the central cell, low-energy ions, that would be confined in absence of the potential, are now pushed away from the central cell by the potential and lost, causing the well-known anisotropic ``ambipolar hole'' in velocity space.

\subsubsection {Axial particle confinement time}

In a simple magnetic mirror, ions are magnetically confined, resulting in an ion confinement time that is typically somewhat longer than the ion-ion collision time $\tau_{ii}$. In 1961, Bing and Roberts \cite{Bing:1961} showed theoretically, employing the Fokker-Planck equation, that the ion confinement time $\tau_i$ can be approximated by
\begin{equation}
	\tau_{i} \approx 2 \log_{10}(R_M) \tau_{ii} = 0.87 \ln (R_M) \tau_{ii},
\end{equation}

\noindent where $R_M$ is the mirror ratio (see also \cite{Fowler:1981} and \cite{Post:1987}). 

Electrons are additionally confined by the positive electrostatic, ambipolar potential $\Delta \phi$. In 1974, Pastukhov \cite{Pastukhov:1974} derived a relation for the axial confinement time of electrons $\tau_e$ in a simple mirror, also employing the Fokker-Planck equation;

\begin{equation}
	\tau_e = \tau_{ee} \ g(R_M)(e \Delta \phi / {T_{e}}) e^{e\Delta \phi / {T_{e}}},
\end{equation}
\noindent where, for $R_M \gg 1$,
\begin{equation}
	g(R_M) = \sqrt \pi \hspace{0.25em} \frac{2R_M + 1} {4R_M} \ln (4R_M+2) \hspace{0.25em} .
\end{equation}

Some basic assumptions of this analysis are: 1) the ambipolar electrostatic potential is so large that the electron population is only marginally perturbed in phase space by losses and remains Maxwellian, so that the Fokker-Planck equation can be linearized, 2) collisions drive electrons into the loss cone, 3) the magnetic field and electrostatic potential are modelled by square wells and 4) the Rosenbluth potentials are valid in the appropriate limits. 

We can now estimate the ambipolar potential $\Delta \phi$, since maintaining quasi-neutrality in steady state requires that
\begin{equation}
	\tau_{e} = \tau_{i}.
\end{equation}

Employing Eqs. (10)-(13), and the relations (1) and (2) for $\tau_{ee}$ and $\tau_{ii}$ we find, for a 50-50 DT plasma with $R_M=10$ and $T_i/T_e=10$, $4$ and $1$ respectively, the values $C_\phi = e\Delta\phi/T_e =5.8$, $4.6$ and $3.0$. Note that the ambipolar potential is completely specified by the two parameters $R_M$ and $T_i/T_e$ alone. The dependence on $R_M$ is however weak. Now we can, using Eq. (7), compute the simple mirror relative density in the mirror throats; they are $n_M/n_0=0.0033, 0.010$ and $0.053$ for $R_M=10$ and $T_i/T_e=10$, $4$ and $1$ respectively.

The loss rate of Eq. (10) yields, as shown in Section 2.3.1, values of the gain factor $Q \lesssim1$, too small to be of interest for a power plant.

\subsection{Classical Novatron mirror, ambipolar state}
An early Novatron design, a simple magnetic mirror with ring-shaped rather than point mirrors, is presented in Figure 1. The axisymmetric vacuum magnetic field, providing plasma interchange stability \cite {Conceptpaper2024}, is shown. The bulk plasma particles bounce between the mirrors, whereas the small fraction of electrons and ions passing near the two magnetic field null regions will experience non-adiabatic excursions, partially into the upper and lower ``chimneys'', as discussed in Section 3. 

\begin{figure}[h!]
	\centering
	\includegraphics[width=2.5in]{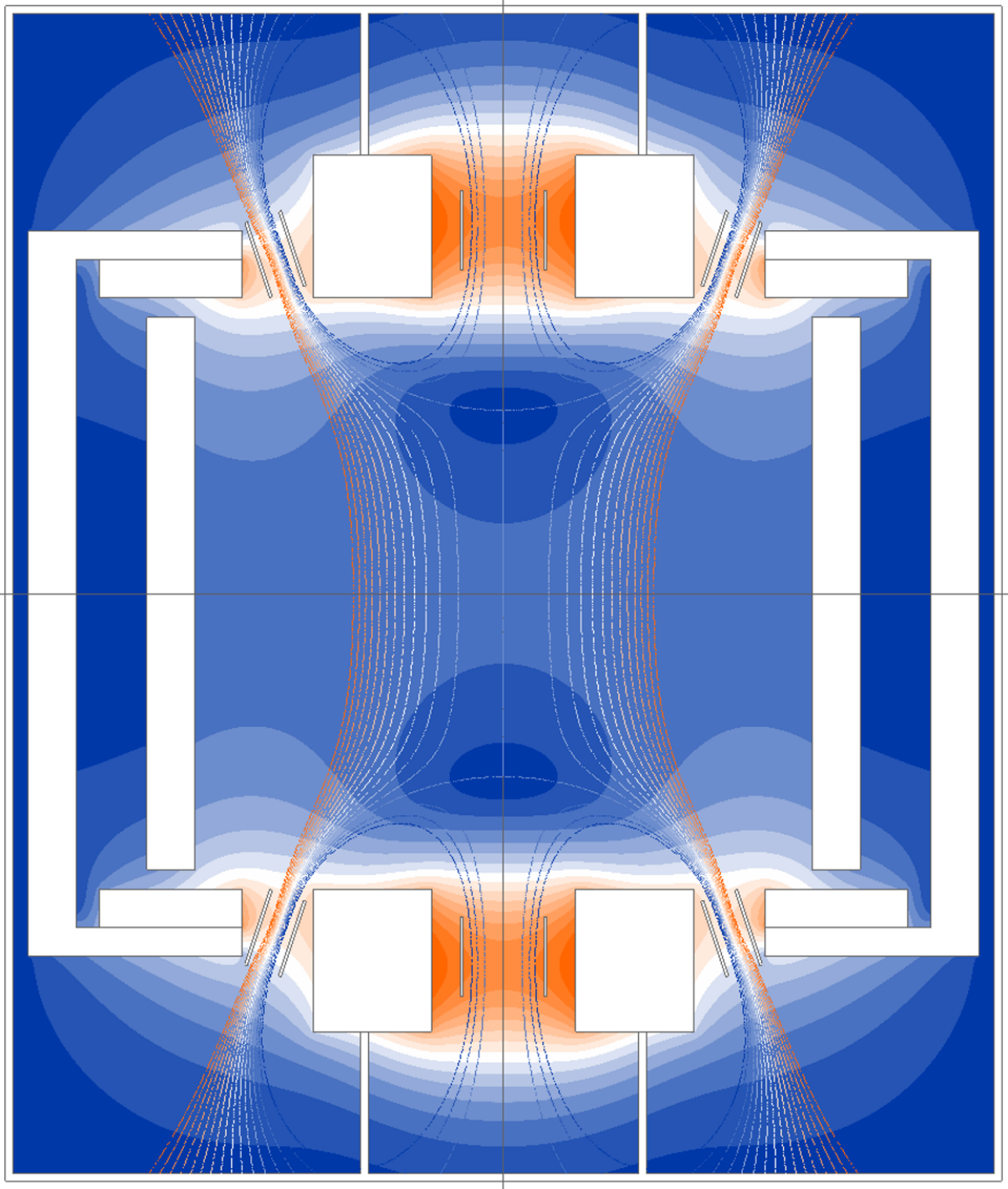}
	\caption{Novatron simple mirror machine design, similar to the N1 experiment featuring a 3.8 m tall, 1.8 m wide central cell including expanders. Magnets are indicated as white sections. Magnetic field lines passing through the magnetic mirrors and the central ``chimneys'' are also shown. The iso-$B$ surfaces are colour-marked from weak field (dark blue) to strong field (orange). Electrodes for investigating ponderomotive RF plugging are indicated in mirrors and chimneys.}
\end{figure}

In Figure 2, a typical Hybrid-PIC simulation is shown. A Hybrid-PIC code in plasma physics \cite{fedeli_pushing_2022,groenewald_accelerated_2023} describes the electrons as a fluid and the ions as kinetic particles. The  computational domain $\Omega = \{ (x, y, z) \in \mathbb{R}^3 \mid -0.7 \leq x \leq 0.7, -0.7 \leq y \leq 0.7, -1 \leq z \leq 1 \}$  was discretized into $96\times96\times96$ cells. The initial conditions were computed from a magneto-static anisotropic equilibrium profile. The initial temperatures were set to $T_e=T_i=100$ eV and the density floor ratio was set to $n_f/n_0=10 \%$, where $n_f$ is the floor density in the vacuum region, and $n_0$ is a reference density set to the maximum number density of the initial condition. Additional parameters include time-step $\Delta t=5\times10^{-10}$ s, resistivity $\eta=10^{-6}$ $\text{V}\text{m} / \text{A}$, hyper-resistivity $\eta_H=10^{-6}$ $\text{V}\text{m}^3 / \text{A}$, $2\times10^{8}$ macro-particles, first-order Direct particle shape, 1-pass filter, and ion-ion collisions.

This simulation was run for $t=100\ \mu s \approx 32.9 t_A$, where $t_A$ is the Alfv\'en time. The results display that in the absence of electrostatic or ponderomotive chimney plugging, the plasma tends to develop into an annular configuration, for which the plasma particles almost exclusively bounce between the ring-shaped mirrors alone, and more specifically in regions where the magnetic moment is conserved.
\begin{figure}[h!]
	\centering
	\includegraphics[width=2.5in]{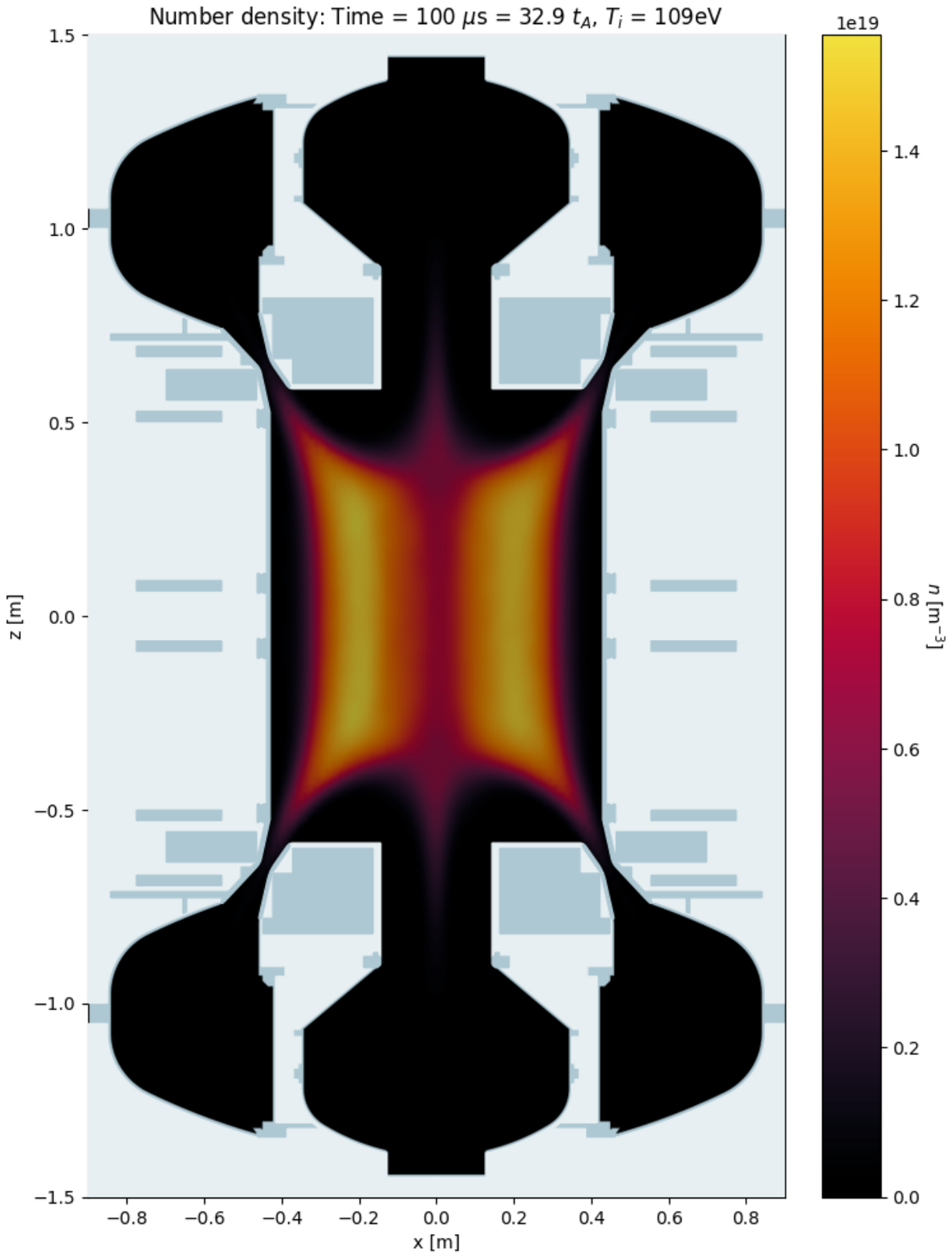}
	\caption{Hybrid-PIC \href{https://ecp-warpx.github.io/}{WarpX} simulation of plasma equilibrium in a simple Novatron configuration. The density surfaces indicate an annular plasma shape, with field lines reaching primarily towards the ring-shaped mirrors. For details, see Section 2.3.}
\end{figure}

Clearly, there are essential similarities between axial confinement, and resulting $Q$-values, in traditional magnetic mirrors and in simple Novatron ring-shaped mirror systems. 

\subsubsection{Q-value}
We wish to estimate theoretically attainable scientific $Q$-values for both simple classical and simple Novatron mirror machines. For steady state, the standard definition is 
\begin{equation}
	Q \equiv \frac {P_{fus}}{P_{in}} = \frac{g_{12}n^2 <\sigma v >_{12} E_{12}}{3nT / \tau_E},
\end{equation}
\noindent where $n$ is the density, $<\sigma v >_{12}$ is the reactivity between species $1$ and $2$ (D-T, D-He3, or D-D),  $g_{12}=1/4$ for the D-T and D-He$^3$ reactions and $1/2$ for D-D, $E_{12}$ is the energy produced per reaction, and $\tau_E$ is the energy confinement time. For a simple mirror, we assume that
\begin{equation}
	\tau_E \approx \tau_i = 0.87 \ln (R_M) \tau_{ii}.
\end{equation}
\noindent A generalization of Eq. (2) is
\begin{equation}
	\tau_{ii} = 2.1 \cdot 10^{13} \frac {T_i^{3/2}} {n Z^2 \ln \Lambda} \sqrt \frac {m_{12}}{m_p} 
\end{equation}
\noindent where the ion temperature $T_i$ here is in units of eV. Also, $m_{12}$ and $m_p$ stand for the average mass of species 1 and 2, and the proton mass. For two ion species with different charges, we assume equal densities and define $Z^{-2} =(Z_1^{-2} + Z_2^{-2})/2$. Combining Eqs. (14)-(16), we obtain
\begin{equation}
	Q = 3.2 \cdot 10^{17} \ln (R_M) \frac {T_i^{1.5}}{T_{ei}} \sqrt \frac {m_{12}}{m_p} \frac {g_{12}}{Z^2} <\sigma v >_{12} E_{12},
\end{equation}

\noindent where $E_{12}$ is now in MeV, $T_{ei} \equiv (T_e+T_i)/2$ in units of eV and $\ln \Lambda =19$ is taken. Let us compare attainable Q-values for the D-T, D-He$^3$, and D-D reactions including catalyzed D-D, where the reaction products T and He$^3$ are also allowed to fuse. For these reactions, $E_{12}=17.6, 18.3, 3.65$ and $21.5$ MeV, respectively. Assuming $R_M = 10$, the following values are  obtained for $T=20$ keV and $T=50$ keV, respectively
\begin{equation}
	\begin{aligned}
	D-T: Q &= 0.30,\ 1.0 \\
	D-He^3: Q &= 1.8 \cdot 10^{-3},\ 0.040  \\
	D-D: Q &= 1.4 \cdot 10^{-3},\ 9.0 \cdot 10^{-3}  \\
	D-D, \text{cat}: Q &= 8.0 \cdot 10^{-3},\ 0.053  \\
	\end{aligned}
\end{equation}

It should be noted that when calculating the reaction energy for the D-D reaction we have employed an average of the two reactions with reaction products He$^3$+n or T+p, yielding 3.65 MeV. Permitting subsequent D-T reactions (``catalyzed DD cycle'') adds 17.6 MeV to the energy 4.03 MeV generated in 50 percent of the reactions. By instead removing a substantial fraction of the tritons \cite{Kesner:2004} before they fuse and replacing these with the He$^3$ decay products (``helium catalyzed D-D cycle'') a total of 22.3 MeV is generated for 50 percent of the D-D reactions, without generating fast, penetrating 14.1 MeV neutrons. 

It is obvious that single mirror particle confinement is insufficient for power plant conditions. This holds for both the simple classical and the simple Novatron mirror machines. Fortunately, electrostatic and RF plugging, realized through the ponderomotive force, may provide a crucial contribution to confinement, as explored in Section 5. 

\section{Adiabatic particle motion in the Novatron}
It can be seen from Figure 1 that the Novatron vacuum magnetic field features two low-B regions, centered on the $z$ axis in the centre cell plasma region. Particles passing near these x-points will become non-adiabatic, implying that our equations for $v_\parallel$ are not valid for these. After a number of passages through the x-point regions it is expected that the particles will spread into the loss cone. This is also confirmed by Hybrid-PIC simulations \cite{fedeli_pushing_2022,groenewald_accelerated_2023} which, for the simple Novatron, show an evolution towards an equilibrium with annular shape, where the region near the $z$-axis is essentially void of particles. MHD stability of the inner region may however still be upheld \cite{Conceptpaper2024}. 

It has been shown \cite {Conceptpaper2024} that for a classical Novatron mirror in low-$\beta$ conditions, the non-adiabaticity of the magnetic moment $\mu$ is limited to a fairly small region close to the central axis. Outside of this region, $\mu$ is conserved to a similar extent as for a simple mirror magnetic field. Since the ponderomotive and ambipolar forces, to a first-order approximation, act to mirror the particle motion along the magnetic field lines, the pitch angle relative to the field line would not change, and thus $\mu$ would remain constant. Consequently, the introduction of these plugging mechanisms is not expected to significantly affect the conservation of $\mu$, nor the leakage into the non-adiabatic region. The dynamics of leakage into and out of the non-adiabatic regions will be a focus of study in the N1 experiment. 

It may be noted that the randomization and isotropization in velocity space associated with non-adiabaticity may act beneficially with respect to micro-instabilities, in particular DCLC modes.

\section{Q enhancement}
The need for improving Q, beyond the values obtainable in the simple mirror systems which were created in the early 1970s in the USA and in Japan, led to the development of more advanced confinement systems and technologies. Tandem mirror systems for Q enhancement were suggested \cite{Sato:1985,Fowler:1977,Dimov:1976,Baldwin:1979} and experiments like TMX, TMX-U and MFTF in the USA, and GAMMA 6 and GAMMA 10 in Japan, were initiated. As discussed further on, the Q enhancement measures taken were, however, not enough to compete with the successful tokamak. Nor did they extrapolate to an operational fusion power plant.

The Novatron demonstrates two significant advantages over other devices in the context of Q enhancement. First, the vacuum magnetic field features favourable curvature so that the plasma is interchange stable; there is no need, like in earlier mirror designs, to find ways to ``anchor'' the unfavourable curvature magnetic field lines using external, favourable curvature structures. Secondly, the device, and its magnetic field, is perfectly symmetric around the $z$-axis. This strongly reduces neoclassical radial transport, supporting the idea that high Q values should be attained by reducing axial particle transport alone.   

In collaboration with mirror fusion specialists K. Fowler and A. Molvik \cite{private_comm}, a comprehensive compilation of nearly 20 potential measures for Q enhancement in Novatron tandem geometry has been developed. The following measures were identified to have the greatest potential impact in a Novatron setting: high plasma temperature, high mirror ratio, electrostatic end plugging, ponderomotive (RF) plugging, beta enhancement, alpha particle confinement and direct conversion of lost particle energy to electricity.

High plasma temperature is beneficial for two reasons. The first is that the fusion reactivity increases with temperature. The second is that the ion-ion collision time $\tau_{ii}$ increases with temperature and consequently less particles are spread into the loss cone. The effect, however, is reduced by the fact that the plasma kinetic energy, in the denominator of Eq. (14), scales as $T$. 

A higher mirror ratio $R_M$ results in, as seen from the equations for $v_\parallel$, a larger fraction of the particles, coming from the central plasma, bouncing at the magnetic mirrors.

Electrostatic and ponderomotive plugging for tandem geometry are discussed in Section 7. High ambipolar electrostatic fields can be obtained in the tandem (plug) cells by increasing the density in these cells or, better, by electron cyclotron heating (ECH) or neutral beam heating (NBI) of the plug cell electrons to a temperature above that of the central cell electrons. Ponderomotive plugging can be used in the tandem cells to contain the heated plug cell electrons, but also for reducing the influx of colder electrons from the central cell. A near 10-fold increase in confinement time was achieved by RF plugging of ions in the RFC-XX device \cite{Sato:1986}.

Beta enhancement can affect Q through enhanced particle confinement time. Higher beta results in a deeper magnetic well, with accompanying enhanced mirror ratio $R_M$ and consequently magnetic particle confinement.

Whereas a mirror fusion reactor can be driven, alpha particle confinement and heating is important for optimizing performance. In particular, magnet and wall structures must be sufficiently isolated from the plasma due to the large Larmor radii of the high energy alpha particles.

Direct conversion of lost particle energy to electricity is attained by implementing suitable schemes, like venetian blinds \cite{Fowler:2017}, in connection to the open field lines emanating from the tandem cells. Particle kinetic energy to electricity conversion efficiencies of 70-80 percent may be expected.

In summary, to substantially improve on the Q-values in Eq. (18), expected for simple mirror geometry, powerful Q enhancement measures must be taken. As shown in Section 6, the most powerful ones are electrostatic and ponderomotive plugging at high central cell plasma temperature. These are to be employed in the Novatron N2 device. Furthermore, a Q enhancement factor of about 3 is expected from direct conversion. Additionally, scenarios with beta enhancement and high mirror ratios ($>10$) will be investigated.

\section{Simple mirror, ambipolar + RF plugging}
The ponderomotive force, as explained in Appendix B, can be written in potential form as $\textbf F_p = - \nabla \Psi$, where $\Psi$ is the ponderomotive potential. Combining the ponderomotive with the electrostatic force, we obtain an extended version of Eq. (9):
\begin{equation}
	v_{\alpha \parallel} = v_{\alpha 0} \sqrt {1-\frac{B}{B_0} \sin^2 \nu + [q_\alpha (\phi_0 - \phi) - \Psi_\alpha]/(m_\alpha v_{\alpha 0}^2/2)}  
\end{equation}

\noindent where the ponderomotive potential has the general form
\begin{equation}
	\Psi_\alpha = \frac{q_\alpha ^2}{4m_\alpha}\Bigg(\frac{E_{\Psi \parallel}^2-E_{\Psi \parallel0}^2}{\omega^2} + \frac{E_{\Psi \perp}^2-E_{\Psi \perp0}^2}{\omega^2-\omega_c^2}\Bigg) .
\end{equation}

\noindent Here $E_{\Psi\parallel}$ and $E_{\Psi\perp}$ are the components of an externally applied RF electric field, with frequency $\omega$ and reference point values $E_{\Psi \parallel 0}$ and $E_{\Psi \perp 0}$. The cyclotron frequency is denoted by $\omega_c$. Since RF plugging is locally applied at the mirror throats (using single or multiple parallel electrodes) we may set $E_{\Psi \parallel 0} = E_{\Psi \perp 0} = 0$. It is seen from Eq. (19) that RF plugging decreases $v_{\parallel}$ for all particles, promoting particle bouncing if acting physically near the mirrors.  

The ponderomotive force can repel both electrons and ions simultaneously, but may act differently on the two species due to the strong inverse mass dependence, and also due to the choice of frequency $\omega$ and direction of the RF field. For reference, we write the ponderomotive potentials for electrons and ions, assuming $B=B_M$, as 
\begin{equation}
	\Psi_{e} = \frac{\omega_{cMe}^2}{{\omega}^2}m_e\frac{E_{\Psi M\parallel}^2}{4B^2_M} + \frac{\omega_{cMe}^2}{{(\omega}^2-\omega_{cMe}^2)}m_e \frac{E_{\Psi M\perp}^2}{4B^2_M}  
\end{equation}

\begin{equation}
	\Psi_{i} = \frac{\omega_{cMi}^2}{{\omega}^2}m_i\frac{E_{\Psi M\parallel}^2}{4B^2_M}  + \frac{\omega_{cMi}^2}{{(\omega}^2-\omega_{cMi}^2)}m_i \frac{E_{\Psi M\perp}^2}{4B^2_M}  
\end{equation}

\noindent where $\omega_{cMe}$, $\omega_{cMi}$ and $E_{\Psi M}$ are the electron and ion cyclotron frequencies and electric fields, respectively, in the region where RF is applied.

In the RFC-XX experiments at Nagoya \cite{Fujita:1988}, where ions were RF-plugged at a frequency just above the ion cyclotron frequency, the electric field was applied by use of long parallel plates. Ratios $E_{\Psi \parallel}/E_{\Psi \perp} = 2 \cdot 10^{-3} - 1 \cdot 10^{-2}$ were found from estimates of the wave numbers of the excited ion Bernstein waves, and the first term of Eq. (22) was negligible. It is of crucial importance to avoid $\omega \approx \omega_c$ resonances (where Eqs. (21) and (22) do not apply), due to the associated particle heating and loss of RF power; see also Appendix B.

For the Novatron, the ponderomotive plugging of both central cell electrons and ions, as well as tandem cell electrons, are of interest. Let us first study the requirements for RF plugging of the central cell so that all particles with speed $v_0$ and pitch angle $\nu$ will be confined. We rewrite the electrostatic term in Eq. (19) using the Boltzmann relation for constant $T_e$;
\begin{equation}
	e\phi = T_e \ln (n) + C,
\end{equation}

\noindent where $C$ is a constant, so that we have 
\begin{equation}
	e\phi - T_e \ln (n) = e\phi_0 - T_e \ln (n_0) 
\end{equation}
\begin{equation}
	e(\phi_0 - \phi) = T_e \ln (n_0/n) .
\end{equation}

\noindent Subscript $``0"$ denotes values taken at the midplane ($z=0$) of the center plasma cell. It should be noted that Eq. (23) is likely to be somewhat modified close to the RF plugging electrodes, but we here assume that the Boltzmann relation is upheld by the fast electrons and that charge separation effects due to ponderomotive forces are small. It is helpful to write Eq. (19) as separate equations for electrons and singly charged ions:
\begin{equation}
	v_{e \parallel} = v_{e0} \sqrt {1-\frac{B}{B_0} \sin^2 \nu - (T_e \ln (n_0/n) + \Psi_e)/(m_ev_{e0}^2/2)} 
\end{equation}
\begin{equation}
	v_{i \parallel} = v_{i0} \sqrt {1-\frac{B}{B_0} \sin^2 \nu + (T_e \ln (n_0/n) - \Psi_i)/(m_iv_{i0}^2/2)}
\end{equation}

In order to obtain $v_{e\parallel} = v_{i\parallel} = 0$ at the mirrors, \textit{the ion equation} (27) poses the most severe criterion due to the repulsive electrostatic force. For $\nu =0$, it must hold that
\begin{equation}
	\frac{\omega_{cMi}^2}{{\omega}^2}m_i\frac{E_{\Psi M\parallel}^2}{4B^2_M}  + \frac{\omega_{cMi}^2}{{(\omega}^2-\omega_{cMi}^2)}m_i \frac{E_{\Psi M\perp}^2}{4B^2_M} = m_iv_{i0}^2/2 +T_e \ln (n_0/n)
\end{equation}

\noindent The ponderomotive force on ions can be optimized by choosing the RF electric field component to be perpendicular to the magnetic field, in which case the resonance near $\omega_{cMi}$ can be utilized. Thus, we here set the first term to zero. Clearly, the \textit{high-energy} ions are the most demanding. Assuming $T_e=T_i=m_i v_{Ti}^2/2$, and setting $v_{i0}=sv_{Ti}$, so that $s$ denotes the ratio between particle velocity and the thermal velocity, we obtain from Eq. (28) the minimum required electric field for ion confinement:

\begin{equation}
	E_{\Psi M\perp}=\sqrt{2(s^2+ln(n_0/n))} \hspace{0.25em} v_{Ti} B_M \sqrt {\frac {\omega^2}{\omega_{cMi}^2}-1} \hspace{0.25em}.
\end{equation}

Furthermore, we make use of the computed value for the ambipolar density ratio $n_0/n$ for the simple mirror, computed in Section 2.2. For $R_M=10$ and $T_i/T_e=1$, we obtained $n_0/n=18.9$. For 20 keV D ions and a magnetic field strength $B_M=10$ T there results, for the simple mirror
\begin{equation}
	E_{\Psi M\perp}=20 \cdot 10^6  \sqrt{s^2+2.9} \hspace{0.25em} \sqrt {\frac {\omega^2}{\omega_{cMi}^2}-1} \quad \text{V/m} .
\end{equation}

In laboratory plasmas DC electric fields up to $10^5\ \text{V/m}$ are reached, and in lightnings $10^6\ \text{V/m}$ have been measured. As considered in the Discussion section, it remains an open question which AC electric field amplitudes may be achieved in a fusion plasma. Presumably Eq. (30) displays a quite narrow band of allowable frequencies, making it difficult to completely confine ions in the central cell by aid of RF plugging alone. This conclusion is also supported by experiments at RFC-XX \cite{Sato:1985}, where it was found that $\Psi_{i}$ decreased with increasing plasma density.

Turning to \textit{electron confinement} in the simple mirror, the equivalent of Eq. (28) is
\begin{equation}
	\frac{\omega_{cMe}^2}{{\omega}^2}m_e\frac{E_{\Psi M\parallel}^2}{4B^2_M}  + \frac{\omega_{cMe}^2}{{(\omega}^2-\omega_{cMe}^2)}m_e \frac{E_{\Psi M\perp}^2}{4B^2_M} = m_ev_{e0}^2/2 -T_e ln (n_0/n) \hspace{0.25em} .
\end{equation}

The confining effect of the ambipolar electrostatic field is apparent. For low frequencies, like $\omega = \omega_{cMi}$, the first left hand term dominates when applying a parallel ponderomotive field. For confinement of electrons in a simple mirror we thus obtain from Eq. (31), with $\omega = \omega_{cMi}$;
\begin{equation}
	E_{\Psi M\parallel}=\sqrt{2(s^2-ln(n_0/n))} \hspace{0.25em} \frac{m_e}{m_i} v_{Te} B_M.
\end{equation}

\noindent We have set $v_{e0}=sv_{Te}$. Hence all electrons, for which $v_{e0} < v_{Te} \sqrt {ln(n_0/n)}$, are electrostatically confined, and ponderomotive confinement is not needed. For the simple mirror, $n_0/n=18.9$ for $T_i/T_e=1$ and $R_M=10$.  Thus only electrons for which $v_{e0}/v_{Te} < 1.71$ are confined at this high electron temperature. Confinement of all electrons, for a 20 keV D plasma with $B_M=10$ T, requires the electric field strength
\begin{equation}
	E_{\Psi M\parallel}= 0.32 \cdot 10^6 \sqrt {s^2-2.9}  \quad \text{V/m} \hspace{0.25em}.
\end{equation}

\noindent In the Discussion, we consider the possibilities of maintaining parallel electric fields at these low frequencies, given the tendency of the highly mobile electrons to short-circuit parallel fields. 

An alternative is to generate perpendicular electric RF fields with frequencies above both the electron cyclotron frequency and the plasma frequency, eliminating the electric field cancellation effect of mobile electrons. The required electric field strength becomes, in analogy with Eq. (29),
\begin{equation}
	E_{\Psi M\perp}=\sqrt{2(s^2-ln(n_0/n))} \hspace{0.25em} v_{Te} B_M \sqrt {\frac {\omega^2}{\omega_{cMe}^2}-1} .
\end{equation}

For a 20 keV D plasma, with $B_M=10$ T, we obtain
\begin{equation}
	E_{\Psi M\perp}=1180 \cdot 10^6 \sqrt{s^2-ln(n_0/n)} \hspace{0.25em} \sqrt {\frac {\omega^2}{\omega_{cMe}^2}-1} \quad \text{V/m} \hspace{0.25em},
\end{equation}

\noindent which is prohibitively high for confining all electrons in the central cell. Frequencies $\omega < \omega_{cMe}$ cannot be employed since in this case the ponderomotive force attracts electrons.

\section{Tandem mirror, ambipolar + RF plugging}

The tandem mirror approach was first suggested by Dimov \cite{Dimov:1976}, then independently by Logan and Fowler \cite{Fowler:1977}. The idea was to enhance ion confinement in the central cell by adding one mirror cell at each end of the device. The plasma potential in these plug cells should then be elevated by either increasing the plug cell density significantly above the central cell density or by heating the plug cell electrons to a higher temperature. The basic physics of this scheme can be understood combining the Boltzmann relations for the central and plug cells. Central cell ions in the loss cone would reflect off of the rising potential in the plugs, that is off the inner edge of the tandem cell plasma. The central cell electrons should partially be electrostatically confined, closer to the midplane on the inner side of the central cell magnets, as well as reflected off of the outer side of the tandem cell plasma, where the plasma is dropping towards ground potential.

It was soon understood that it is difficult to significantly heat the plug cell electrons, due to their collisional mixing with the cooler central cell electrons near the magnetic mirrors. Furthermore, the logarithmic density dependence of the potential would require very high densities in the plug cells ($n_p \approx 10 n_c$) which, in turn, would require very large magnetic fields in the plugs (of order $20$ T), as well as neutral beam heating particle energies of up to 1 MV in a reactor.

Thus, in 1979, a remedy was suggested in the shape of the ``thermal barrier'' scheme \cite{Baldwin:1979}. The idea was to sequentially confine first electrons, then ions, as seen from the central cell. The central cell electrons would be stopped from entering the plug cell by a local dip in the potential between the central cell and the plug cell. This could be accomplished by installing additional mirror magnets between the central cell and the plug cells, and applying ECRF heating to the confined electrons, while pumping out the ions. The TMX-U experiment tried this configuration but it was soon found that the particular phase space architecture required was too demanding. The thermal barrier could not be upheld for central cell densities beyond $10^{18}\ \text{m}^{-3}$ \cite{Post:1987}.

In the present work, we instead suggest the use of the ponderomotive force for separating the central and plug cell electrons, while maintaining a positive plug cell potential for confinement of central cell ions. For this, plug cell electrons, heated above the temperature of the central cell electrons, are confined using RF ponderomotive plugging at the cusp or mirror ends of the Novatron tandem cell region, where the magnetic field lines leave the cells. A tandem Novatron configuration is illustrated in Figure 3. The magnetic topology design establishes distinct domains of confined plasma outside the central cell mirrors. These domains will be utilized for electrostatic plugging of the central cell. The tandem cells are formed by introducing additional pairs of magnets, featuring the same current directions as the Novatron mirror throat magnets. The latter magnets, positioned outside the central cell, creates a classic mirror configuration within a ring-shaped domain. The tandem cell field line shapes and the mirror ratios, towards both the central cell and the expander, are regulated by control magnets, which can have conical or other shapes. This arrangement allows for the control of the ``degree of bad curvature" for the tandem cell at both the radially outer and inner regions. For the case that central cell flux tubes extend into the bad curvature tandem region, the interchange stability integral can be made positive, resulting in MHD stability. For highly efficient plugging between the central cell and the tandem cell, however, the bad curvature of the tandem cell can cause interchange instability and trapped particle modes \cite{Berk1983FastGT,Gerver1989,Kadomtsev1971} in the tandem plasma region. The trapped particle modes in the Tara tandem mirror \cite{Gerver1989}, for example, were identified as a result of separation, at high central cell mirror ratio, between the tandem cell and the stable central cell plasmas. This issue is addressed by incorporating an anchor cell, specifically a so-called Nova-cusp cell featuring only line cusps and no point cusps, positioned outside the tandem cell. Thus, the anchor cell can stabilize the tandem cell regardless of the plugging efficiency of the mirror throats. It should be noted that the anchor cell provides superior stability as compared to expanders. By adding ponderomotive plugs towards the expanders, local plasma density increases, further enhancing stability. Presently also different tandem and anchor cell designs are evaluated theoretically. One family of these Novatron designs features good magnetic field curvature in both central and tandem cell plasmas, rendering anchoring, with associated cusp geometry-caused losses, unnecessary.

Novatron central and tandem cell microinstability will, as mentioned in the Introduction, be discussed elsewhere. It may, however, be said that DCLC stability of the central and tandem cell plasmas is upheld by diminishing the hole in ion phase space by electrostatic and ponderomotive plugging as well as by a large ratio of plasma to Larmor radius \cite{Simonen1978}. Sloshing ion stabilization, created by inclined NBI \cite{Kotelnikov2017}, does not seem to be necessary.

In a tandem mirror, it is of interest to determine the effect of the electrostatic field, $\Delta \phi = \phi_p - \phi_c$, on ion confinement, where $\phi_p$ is the tandem (plug) cell potential and $\phi_c$ is the central cell potential. The Pastukhov analysis presented in Section 2, however, only holds for electrons. Thus, in 1978, Cohen et al \cite{Cohen:1978}, generalized Pastukhov's result to confinement of arbitrary particle species, including multiple-species with multiple charges. They also generalized Pastukhov's square well magnetic field to arbitrary magnetic-field profiles. Their expressions for axial electron and ion confinement times are as follows:

\begin{equation}
	\begin{split}
	\tau_e = \tau_{ee} \frac {\sqrt \pi} {4} \frac{2R_M + 1} {2R_M} \ln(4R_M+2) (e (\Delta \phi)_e / {T_{e}}) e^{e (\Delta \phi)_e / {T_{e}}} \\
	\tau_i = \tau_{ii} \frac {\sqrt \pi} {2} \frac{R_M + 1} {R_M} \ln(2R_M+2) (e (\Delta \phi)_i / {T_{i}}) e^{e (\Delta \phi)_i / {T_{i}}} \\
	\end{split}
\end{equation}

The relations hold for singly charged, single ion species plasmas. It should be noted that $(\Delta \phi)_e$ and $(\Delta \phi)_i$ in these expressions denote the \textit{confining} potential difference for electrons and ions, respectively. For example, the relation for $\tau_i$ in Eq. (36) is not valid in a simple mirror, since the confining, positive ambipolar potential is that of the electrons. This holds since these relations are derived assuming that the electrostatically confined species features a near-Maxwellian distribution, due to that $e \Delta \phi/T_\alpha \gg 1$, so that the Fokker-Planck equation can be linearized. It may also be noted that the equation for $\tau_e$ includes a correction by a factor 2 of Pastukhov's result. The corrected formula leads to the following more accurate values of the ambipolar potential in the simple mirror for $R_M=10$ and $T_i/T_e=10$, $4$ and $1$ respectively (see Section 2.2.2); $C_\phi = e\Delta \phi/T_e=6.4$, $5.2$ and $3.5$. 

An interesting observation by Cohen et al is that electrons pitch-angle-scatter off of \textit{both} ions and electrons, with similar collision times for singly charged ions, whereas the ions scatter nearly exclusively off of only ions, thus resulting in
\begin{equation}
	\frac {dn_e/dt}{dn_i/dt} \approx 2  ,
\end{equation}

\noindent that is the density deterioration for electrons, in relation to that of ions in similar plasmas (but reversed confining potentials), is a factor of 2 faster.

Cohen et al \cite{Cohen:1978} compared their analytical results with Fokker-Planck code simulations and found an error of only 10-20 \%. Even somewhat higher accuracy for their analytic estimates of electron and ion confinement times was obtained in 1984 by Najmabadi et al \cite{Najmabadi:1984}. For cases with frequent collisions, so that the mean free path is reduced to the order of the mirror system length, Rognlien and Cutler \cite{Rognlien:1980} in 1980 published a Fokker-Planck study, showing that an approximate expression for the composite confinement time is $\tau = \tau_c + \tau_P$, where $\tau_c$ is the collisional time and $\tau_P$ is a Pastukhov confinement time. Thus the model results in smooth transitions for the confinement time when passing from the collisional to the collisionless regime. 

\begin{figure}[h!]
\includegraphics[width=0.8\linewidth]{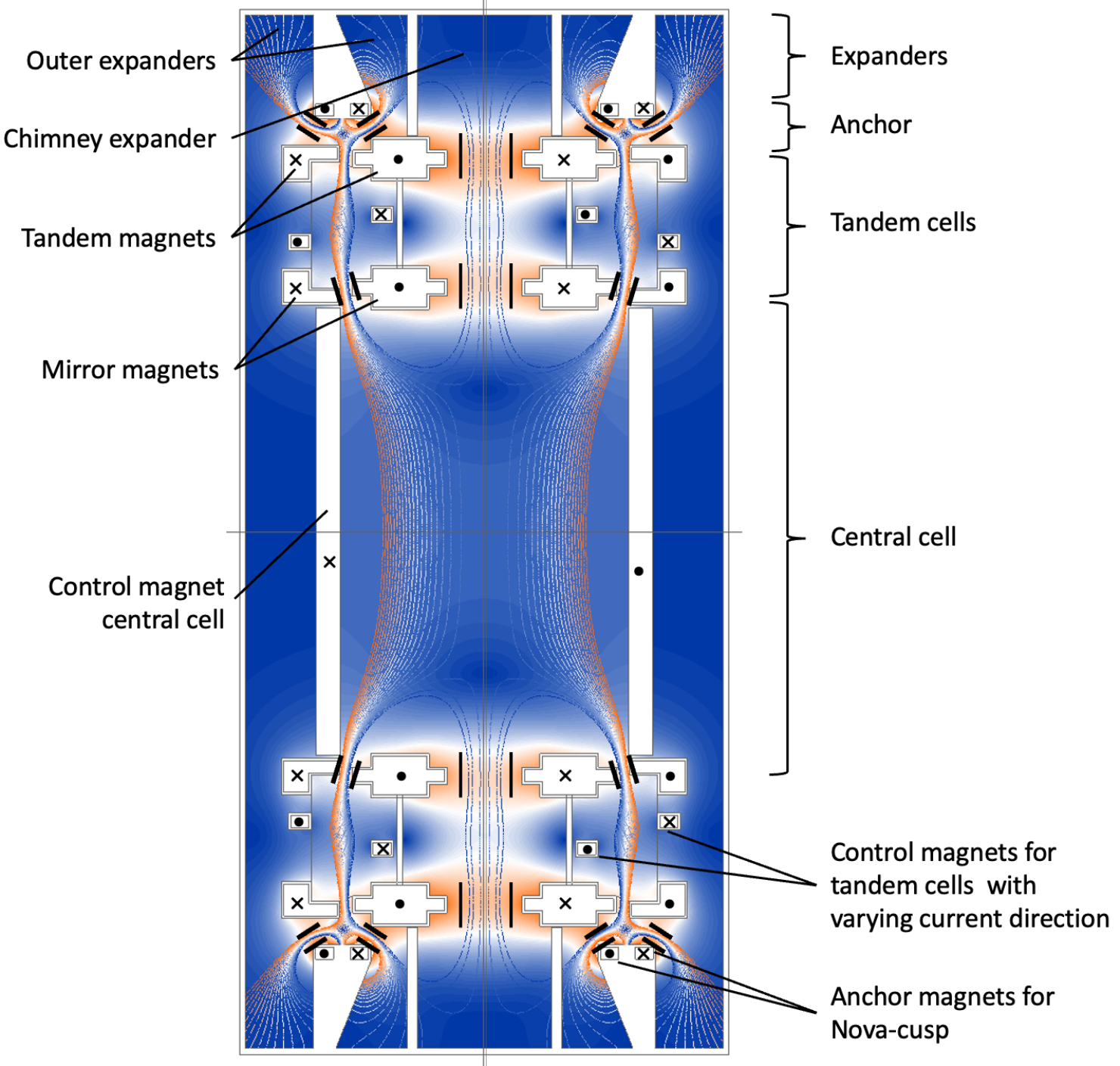}\hfill
\includegraphics[width=0.2\linewidth]{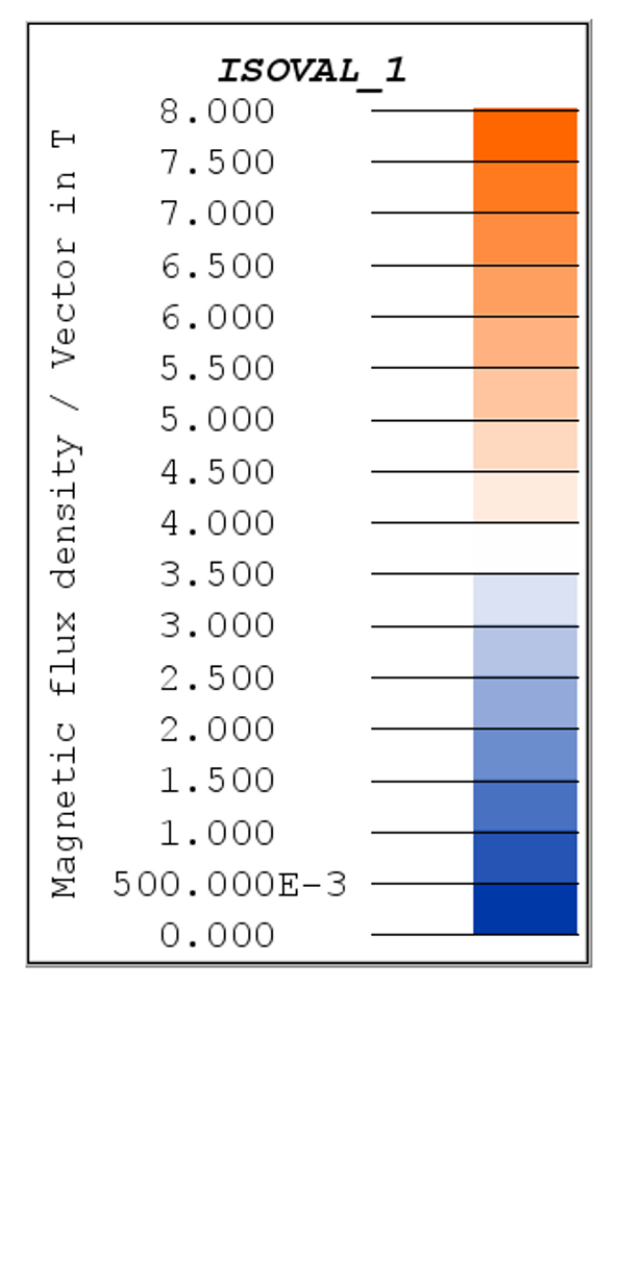}
\caption{A tandem Novatron mirror machine, as presently envisaged. The scale of the figure is $4 \times 8.5$ m. Central, tandem (plug) and cusp expander cell magnets and field lines are shown. Electrodes for RF plugging of tandem and expander cells, as well as chimneys (near symmetry axis) are also indicated. The coding for magnetic field strength is provided on the right.}
\end{figure}

\subsection {Enhanced Q in Novatron tandem geometry}

In a tandem Novatron mirror, central cell ions are mainly confined by the positive electrostatic field of the  plug cell. Additionally, as indicated by Eq. (31), RF plugging of tandem cell electrons also appears possible. It should be noted that complete electron confinement, as investigated in the previous section, is not required. The aim is to, with reasonable external heating power, heat and confine plug cell electrons, maintained at a temperature above that of the central cell electrons, as derived below. This provides the elevated, positive plug potential needed to confine central cell ions. 

One should bear in mind, however, that ponderomotive plugging of the mirror throats to isolate the colder central cell electrons from the plug cell electrons is probably necessary, as learned from the thermal barrier experiments in the US. While complete RF plugging of central cell ions in the simple mirror configuration seems overly demanding, partial electron plugging in a tandem configuration should be more achievable. This can be accomplished by generating suitable RF electric fields with parallel components, a technique that will be experimentally investigated in the Novatron N1 device.

We now turn to study scenarios for elevating the positive electrostatic potential of the tandem cells to levels sufficient for reactor relevant central cell ion confinement times. We also discuss how ponderomotive plugging may aid in this process. Generalizing the Boltzmann equation to take into account the temperature difference $T_{ep}-T_{ec}$ between the plug and central cells, we integrate $ed\phi/dz = (T_e/n)dn/dz + dT_e/dz$ across the mirror region, where the density is assumed to have a local minimum $n=n_M$, to obtain
\begin{equation}
	e\Delta \phi = e\phi_p - e\phi_c = T_{ep} (1+\ln(n_{p0}/n_M)) -T_{ec} (1+\ln(n_{c0}/n_M)), 
\end{equation}
\noindent where index ``0" denotes maximum value in the cell, resulting in

\begin{equation}
	e\Delta \phi/T_{ec} = \frac{T_{ep}}{T_{ec}} -1 + \ln\big((n_{p0}/n_M)^{T_{ep}/T_{ec}}/(n_{c0}/n_M)\big).
\end{equation}

The Cohen (1978) formula for the ion confinement time is now generalized to include the effect of a ponderomotive potential $\Psi_{pi}$ acting on ions;

\begin{equation}
	\tau_i = \tau_{ii} \frac {\sqrt \pi} {2} \frac{R_M + 1} {R_M} \ln(2R_M+2) \frac {e\Delta \phi + \Psi_{pi}} {T_{i}} e^{(e \Delta \phi +\Psi_{pi})/ {T_{i}}} 
\end{equation}

This formula is motivated by the fact that ions approaching the mirrors, or chimneys, where a ponderomotive force is generated, cannot distinguish its effect from that of the electrostatic force of the tandem cell plasma. The strong confining effect of the combined plug electrostatic potential and ponderomotive force can be seen by assuming $R_M=10$ and setting $(e\Delta \phi + \Psi_{pi})/T_{i}  = 2.5$, obtaining
\begin{equation}
	\tau_i = 92 \tau_{ii} .
\end{equation}

For $(e\Delta \phi + \Psi_{pi})/T_{i}  =5$, we obtain
\begin{equation}
	\tau_i = 2.2 \cdot 10^3 \tau_{ii} .
\end{equation}

In the latter case, the simple mirror central cell Q-values (18) are now replaced with (for T = 20 keV and T = 50 keV, respectively):
\begin{equation}
	\begin{aligned}
	D-T: Q &= 3.4 \cdot 10^2,\ 1.1 \cdot 10^3 \\
	D-He^3: Q &= 2.0,\ 45  \\
	D-D: Q &= 1.6,\ 10  \\
	D-D, \text{cat}: Q &= 9.3,\ 60  \\
	\end{aligned}
\end{equation}

The effect of electrostatic and ponderomotive tandem cell plugging is obviously strong already at realistic potential magnitudes. Also considering that at least a factor 3 is expected to contribute from additional Q enhancement measures, it is clear that fusion relevant Q-values would be obtainable for the D-T reaction (T = 20 and 50 keV), the D-He$^3$ reaction (50 keV), the D-D reaction (50 keV), and the catalyzed D-D reaction (T = 20 and 50 keV).

The Q-values in (43) are only attainable if sufficient electrostatic and ponderomotive confinement of the central cell plasma can be reached. To iterate, according to the Boltzmann relation, this entails either a strong enhancement of the plug cell density in relation to the central cell density or an elevated plug cell electron temperature. Again, the first alternative was tried in the TMX device, but it was soon found that because of the very high plug cell density, extremely high magnetic fields (order 20 T or more) for plug cell plasma equilibrium and neutral beam energies of order 1 MeV would be needed for reactor conditions. The question is thus whether it is feasible to raise the plug cell electron temperature sufficiently to attain $(e\Delta \phi + \Psi_{pi})/T_{i} \approx 5$.

First, neglecting ponderomotive effects, we note that in a steady-state fusion plasma, the energy equipartition time $\tau_{eq} \approx 1$ s for $T=10-50$ keV. Consequently, we have $T_{ec} \approx T_{ic}$. We can then set
\begin{equation}
	\Delta \phi/T_{ec} = \Delta \phi/T_{ic}
\end{equation}

\noindent and solve Eq. (39) for $T_{ep}/T_{ec}$, assuming $n_{p0} = n_{c0}$. We find

\begin{equation}
	T_{ep}/T_{ec} = 1 + \frac{e\Delta \phi / T_{ic}}{1+\ln (n_{p0}/n_M)}
\end{equation}

Hence, for maximum plug density to mirror density ratios $n_{p0}/n_M = 10$ and $100$, respectively, employing $e\Delta \phi/T_{ic} = 5$ we find:
\begin{equation}
	T_{ep}/T_{ec} \approx 2.5, 1.9  .
\end{equation}

Thus, an increase in Q by about a factor 1000 is obtained by approximately doubling the plug cell electron temperature compared to that of the central cell. The strong sensitivity derives from a two-fold dependence;  first, the ion confinement time in Pastukhov-like formulas for confined ions are both linearly and exponentially depending on the ratio $e\Delta \phi/T_{ec} = e\Delta \phi/T_{ic}$, which in turn, as seen from Eq. (45), scales as $(T_{ep}/T_{ec}-1)\ln (n_{p0}/n_M)$. It may also be noted that since the central cell ion confinement time depends on the sum $(e\Delta \phi + \Psi_{pi})/T_{i}$, a superimposed ponderomotive potential further strengthens ion confinement.

\subsection{Central cell electron plugging}

In order to maintain the temperature ratios $T_{ep}/T_{ec}$ of Eq. (46), it is desirable to prevent the colder central cell electrons from penetrating into the tandem cells. Parallel or perpendicular RF electric fields may be employed for ponderomotive plugging. Applying parallel RF electric fields of low frequency ($\omega \approx \omega_{cMi}$) to a 20 keV D plasma with $B_M=10$ T, we find the required electric field from Eq. (32);
\begin{equation}
	E_{\Psi M\parallel}= 0.64 \cdot 10^6 \sqrt {s^2-ln(n_0/n)}  \quad V/m \hspace{0.25em}.
\end{equation}

A general result, that follows from Eq. (31), is that if the central cell plasma density $n=n_M$ near the mirror region satisfies $ln(n_0/n_M) \ge s^2$, plugging is not needed. By choosing $s=v_{e0}/v_{Te} \approx \sqrt{T_{ep}/T_{ec}} = \sqrt{1.9} = 1.4$, the population of central cell electrons that pass the ponderomotive barrier will not cool the tandem cell plasma. The numerical value follows from the two relations $v_{e0} = sv_{Tec} = s \sqrt{2T_{ec}/m_e}$ and $v_{e0} \le \sqrt{2T_{ep}/m_e}$. For $s^2=1.9$, the limit $n \le n_{max} = 0.15n_0$ obtains. As discussed earlier, magnetic mirror confinement at high mirror ratio contributes to satisfying $n \le n_{max}$, evidenced by the parallel component $\partial p_\parallel/\partial l = -(p_\perp-p_\parallel) \partial lnB/ \partial l$ of the MHD equilibrium equation, where $\partial l$ is a field line segment. Maintaining $p_\perp > p_\parallel$ implies decreasing density towards the magnetic mirrors, assuming nearly constant temperature. Thus \textit{ECH could play a major role for central cell electron plugging}, a topic that deserves further investigation. Additionally, conservation of the magnetic moment $\mu$ increasingly prevents electrons from reaching the strong-field mirror region. However, the tandem mirror distinguishes itself from the simple mirror in that the positive electrostatic potential of the tandem region may reduce the potential $\Delta \phi$ between the interior of the central cell and the mirror region, thus reducing the $n_0/n$ ratios that obtain in the simple mirror (see Section 2.2.2), as required by the Boltzmann relation. More careful analysis and experimental results are needed to understand this mechanism.  

Parallel ponderomotive central cell electron plugging, with the electric field amplitudes given by Eq. (47), must take into account the field cancelling effect of mobile electrons. The latter effect is mitigated by choosing frequencies $\omega$ above the electron cyclotron and plasma frequencies, but the required electric field amplitudes are likely to become too large. Perpendicular ponderomotive plugging of central cell electrons at frequencies just above the electron cyclotron frequency avoids the problem of mobile electron cancellation. We saw from Eq. (35) that this requires very careful matching of the applied frequency in order to attain a sufficiently high electric field amplitude $E_{\Psi M\perp}$.

In summary, we have identified the following methods to prevent cooler central cell electrons from penetrating into the tandem cell regions. Provided that the central cell density near the mirror throats satisfies $n \le 0.15n_0$, electrostatic confinement of central cell electrons is sufficient to maintain a temperature ratio $T_{ep}/T_{ec} \approx 1.9$, consistent with the reactor relevant $Q$-values of Eq. (43). If additional central cell electron plugging is needed, either of electron cyclotron heating (ECH) of central cell electrons to enhance $p_\perp > p_\parallel$, or ponderomotive plugging can be employed. The latter option separates into use of either parallel or perpendicular RF electric fields, both for which different frequencies may be selected. It is found that in order to avoid field cancelling effects due to mobile electrons, high frequencies and thus high RF electric field amplitudes are preferred, but plugging at lower frequencies should be explored. The physics of ponderomotive electron plugging is complex and will be investigated experimentally in the Novatron N1 device, which is currently nearing completion.

\section{Discussion}
\label{sec:discussion}
As for all mirror fusion concepts, the Novatron will face both axial and radial losses as well as charge exchange and radiation losses. Our focus is here on axial losses, believed to be of central importance. The Novatron will rely on triple-force electron and ion axial confinement, of which magnetic mirror and electrostatic forces are well understood both theoretically and experimentally since mirror experiments in tandem geometries started in the late 1970s. The third confining force, that of ponderomotive RF plugging, is not so well understood. Experiments in RFC-XX and Phaedrus, among others, have shown that careful tailoring of antennas and RF frequencies have significant effects on confinement. In a fusion relevant Novatron the demands on antenna design and electric field strengths are more demanding. Thus, already in the Novatron N1 device basic ponderomotive antenna systems will be tested and evaluated, as a basis for the design of the N2 device which in turn prepares for reactor size N3 and N4.     

Since the Q-values of simple mirror machines, with single mirrors at each end of the central cell, are too low for fusion-relevant plasmas, N2 and subsequent devices will be designed as tandem mirror machines. To maintain a higher electron temperature in the tandem cells compared to the central cell, ponderomotive plugging of electrons in either the central or tandem cells may be employed. However, because the loss of hot electrons from the tandem cells can be compensated for by extending the length of the central cell, it is preferable to plug the electrons in the central cell. For these, frequencies above the electron cyclotron frequency $\omega_{ce}=eB/m_e=1.8 \cdot 10^{11} B$ Hz and the plasma frequency $\omega_{pe}=\sqrt {n e^2/\epsilon_0 m_e}=56 \sqrt{n}$ Hz can be employed. For $B=10$ T and $n=10^{20}\ \text{m}^{-3}$ it holds that $\omega_{ce}=1.8 \cdot 10^{12}$ Hz and $\omega_{pe}=5.6 \cdot 10^{11}$ Hz. It would seem that it is necessary to stay above these frequencies in order to avoid short-circuiting by freely movable electrons and that the RF power goes to electron heating via wave-particle resonances. In a patented Novatron design to be tested in the N1 device, however, short-circuiting at low frequencies (for strong ponderomotive effect) is avoided by employing pairs of RF electrodes, adjusted to be out of phase to generate electric fields with parallel components. Since these fields are curved towards the electrodes, electrons, that must follow the magnetic field lines, are prevented from short-circuiting the electric field in the plasma. 

Also, higher frequencies into the terahertz domain should be avoided due to relativistic effects that may reduce efficiency. Traditional antennas or electrodes should be small and precisely designed at these short wavelengths in order to be efficient. The maximum power that can be radiated to a specific spatial region is also affected at these high frequencies. Finally, the RF power can be difficult to couple to the inhomogeneous plasma due to wave spreading and reflection. All these aspects should be thoroughly studied experimentally before construction of N2, to ensure that the ponderomotive electric field's frequency, strength and deposition meet the requirements  for sufficient particle plugging.

For ponderomotive plugging of ions, large electric AC fields with frequencies above the ion cyclotron frequency can be sustained as excited ion Bernstein waves \cite{Fujita:1988} if their propagation is perpendicular to the field lines, avoiding short-circuiting by fast electrons. To prevent power from being diverted to ion heating and Alfv\'en waves with strong resonance coupling, the frequency should be maintained above the ion gyrofrequency.

The limits for electric field strengths in a fusion plasma depend on the direction of the electric field component. For electric field components along the magnetic field lines, electron runaway phenomena must be considered. The classical expression for the Dreicer field, obtained by balancing electric field and collisional effects, is generally too low when compared to experiments. However, it may be used for estimates. 

For electric RF fields, propagating perpendicular to the magnetic field, a limiting $E_{\perp max}$ can be estimated by assuming that the electric field force should be less than the Larmor force acting on the ions in order to avoid significant effects on ion dynamics, including instabilities. Thus, for singly charged ions, the relation $e E_{\perp max} \approx m_i v_{Ti}^2/r_{Li} = m_i \omega_{ci} v_{Ti}$ holds. This yields $E_{\perp max} = 1.0 \cdot 10^4 B \sqrt{T_i}$ V/m, where $T_i$ is in eV. For $B=10$ T and $T=20$ keV the resulting $E_{\perp max} = 1.4 \cdot 10^7$ Vm$^{-1}$ is consistent with the requirements for the perpendicular ponderomotive field amplitude discussed in this paper.

Finally, the success of RF ponderomotive plugging of electrons and ions depends critically on the design of the RF electrodes / antennas and their interaction with the plasma in the regions near the magnetic mirrors and chimneys. In particular, shaping and choice of insulating materials are crucial to avoid ionization, erosion, sputtering and interference with the magnetic field. These phenomena are to be studied in Novatron N2.

\section{Conclusion}
\label{sec:conclusion}
The Novatron is designed as a novel type of axisymmetric magnetic mirror machine, inherently stable against MHD interchange and kinetic DCLC modes. It uniquely employs simultaneous use of the triple forces; magnetic mirrors, electrostatic potentials and ponderomotive forces in a tandem-like configuration. The fusion potential of the Novatron has been addressed by considering axial electron and ion confinement, as well as the ratio Q between produced fusion power and heating power. 

It is found that substantial $Q$ enhancement beyond the simple mirror is obtained by maintaining a high electrostatic potential in the tandem, or plug, cells through preferential heating of the tandem cell electrons. Ponderomotive plugging of the tandem cells is achieved by applying RF voltage to electrodes placed at the mirror throats of the tandem cells. RF plugging electric fields at the mirror throats of the central cell partially separate the tandem cell and cooler central cell electron populations. Electron cyclotron heating (ECH) of central cell electrons can assist the process by enhancing reflection at the central cell mirrors. 

The unique combination of Novatron plasma geometry with magnetic, electrostatic and ponderomotive plugging offers a mirror fusion concept with high fusion potential. D-D fusion scenarios at 50 keV temperatures and catalyzed D-D scenarios both at 20 and 50 keV are modelled, suggesting possible paths beyond D-T fusion.      \\

\section*{Acknowledgements}
This research used the open-source particle-in-cell code WarpX \url{https://github.com/ECP-WarpX/WarpX}, primarily funded by the US DOE Exascale Computing Project. Primary WarpX contributors are with LBNL, LLNL, CEA-LIDYL, SLAC, DESY, CERN, and TAE Technologies. We acknowledge all WarpX contributors.

\bibliographystyle{elsarticle-num}
\bibliography{mybib}

\section* {Appendix}
\renewcommand{\theequation}{A\arabic{equation}}
\setcounter{equation}{0}

\subsection*{A. Particle parallel velocity}
Parallel particle confinement in mirror systems is conveniently studied using an expression for the particle's velocity along the field lines $v_\parallel$. The equation of motion for a particle with charge $q$, mass $m$ and velocity $\textbf v$ in a magnetic field $\textbf B$, under the influence of electrostatic and ponderomotive forces, is
\begin{equation}
	m \frac {d \textbf v}{dt} = -q \nabla \phi - \nabla \Psi + q \textbf v \times \textbf B
\end{equation}

Here $\phi$ is the electrostatic potential and the ponderomotive potential $\Psi$ can be written as (see derivation in Appendix B)
\begin{equation}
	\Psi \equiv \frac {q^2} {4m} \Bigg(\frac {E_\parallel^2}{\omega^2}+\frac {E_\perp^2}{\omega^2-\omega_c^2}\Bigg),
\end{equation}
where $\omega_c=qB/m$ is the cyclotron frequency. The applied electric ponderomotive RF field has the parallel and perpendicular components $E_\parallel$ and $E_\perp$.

Multiplying the equation of motion with $\textbf v$, using the relation $df = \textbf {ds} \cdot \nabla f$, and integrating over time, noting that the $q \textbf v \times \textbf B$ force performs no work on the particle, we obtain the energy conservation equation

\begin{equation}
	\frac {1}{2} mv_\parallel^2 + \frac {1}{2} mv_\perp^2 + q\phi + \Psi = 
	\frac {1}{2} mv_{\parallel0}^2 + \frac {1}{2} mv_{\perp0}^2 + q\phi_0+ \Psi_0
\end{equation}
Indices ``0'' indicate that these quantities are computed as they cross the midplane $z=0$. Mirror confinement depends crucially on the constancy of the first adiabatic moment $\mu$. Thus, we assume
\begin{equation}
	\mu \equiv \frac {mv_\perp^2}{2B} = \frac {mv_{\perp_0}^2}{2B_0}.
\end{equation}

Employing the notation, where $\nu$ denotes the pitch angle,
\begin{equation}
	\begin{split} 
	v_{\perp 0} =v_0 \sin \nu, \\
	v_{\parallel 0} =v_0 \cos \nu, \\
	\end{split}
\end{equation}
\noindent and Eq. (A4), we can write Eq. (A3)
\begin{equation}
	\frac {1}{2} mv_\parallel^2 = \frac {1}{2} mv_0^2 - \frac {1}{2} mv_0^2 \frac {B}{B_0} \sin^2 \nu  + q(\phi_0-\phi) - (\Psi-\Psi_0),
\end{equation}

\noindent finally obtaining

\begin{equation}
	v_\parallel = v_0 \sqrt{1 - \frac {B}{B_0} \sin^2 \nu + \frac {q(\phi_0-\phi)} {mv_0^2/2} - \frac {\Psi-\Psi_0} {mv_0^2/2}}.
\end{equation}

It is clearly seen from this equation that all particles, irrespective of charge, experience a repelling force by a mirror magnetic field $B = B_M > B_0$. Furthermore, electrons, with $q=-e$, are pulled back as they approach the mirrors from the center cell because of the ambipolar potential $\phi_0 > \phi$. The effect of the ambipolar potential on ions, in particular those with low energies, is the opposite. Hence for ion confinement in a simple mirror, it must hold that $T_e \ll T_i$, since if $T_e \approx T_i$ so that $mv_0^2/2$ is similar for electrons and ions, the requirement $e(\phi_0 - \phi) \gg mv_0^2/2$ for electron confinement would expel a large fraction of the ions from the center cell. Finally, it is clear that the ponderomotive force, if applied near the mirror throats so that $\Psi > \Psi_0$, supports confinement of both electrons and ions. 

\renewcommand{\theequation}{B\arabic{equation}}
\setcounter{equation}{0}

\subsection*{B. The ponderomotive force}
The ponderomotive force on a charged particle in a magnetic field is a nonlinear effect due to the spatial gradient of an oscillating electric (RF) field. As will be seen in the following derivation, when spatial variations are taken into account there is constructive interaction between the field oscillations and the oscillating particle position. Or, to be more precise, this nonlinear force arises due to the action of an oscillating electric field on a charged particle, the position of which itself is a function of the oscillating electric field. The force has the same sign for both electrons and ions, but because of the inverse mass dependence the force is generally much stronger on electrons. Due to a resonance, $\textbf E$-fields oscillating near the ion cyclotron frequency can however provide a strong effect on ions. As shown, by also solving the particle equations of motion for RF frequencies $\omega$ near the cyclotron frequency $\omega_c$, resultant resonant energy absorption calls for extended modelling, including collisional drag forces on the particle. Usage of the formulas for frequencies close to $\omega_c$ should therefore be made with caution.  

In the literature on this subject, there is agreement over the mathematical formulas for the ponderomotive force. There is, however, less agreement regarding their derivation and the physical interpretation of the force. We hope to clarify these issues here. The derivation starts with the equations of motion for an individual particle in magnetic and electric fields. These are subsequently solved, assuming a simple form for the oscillating RF field. As a next step, the ponderomotive force is obtained from the equation of motion by inclusion of the spatial variation of the electric field and the associated induced magnetic field, as obtained from Faraday's law. Averaging over some periods of oscillations, most terms vanish. The result is a compact equation, displaying the dependence of the ponderomotive force on $\nabla E_\parallel^2$ and $\nabla E_\perp^2$, as well as on the imposed frequency and the mass and cyclotron frequency of the particle. The approach is inspired by the derivation by Chen \cite{Chen:1} for the case that the background magnetic field $\textbf B=\textbf 0$. A heuristic derivation may be found in Nishikawa and Wakatani \cite{Nishikawa:1}.

\subsubsection*{Basic equations, oscillatory electric field}
The equation of motion for a single particle in a combined electric and magnetic field, as well as Faraday's law are needed:
\begin{equation}
	m \frac {d \textbf v} {dt} = q(\textbf E + \textbf v \times \textbf B) 
\end{equation}
\begin{equation}
	\nabla \times \textbf E = - \frac {\partial \textbf B} {\partial t}
\end{equation}

\noindent Here $m$ and $q$ are the mass and charge of the particle. Faraday's law is preferred over Ampere's law in this context, as it naturally incorporates the crucial spatial gradient of the electric field and avoids the introduction of the unknown current density. The equation of motion is exact if the fields are evaluated at the position of the particle. We write the electric field as
\begin{equation}
	\textbf E(\textbf r) =  -\nabla \phi + \textbf E_s(\textbf r) \cos \omega t  , \\
\end{equation}

\noindent where we include an electrostatic field $-\nabla \phi$ and  a transverse RF wave with time period $T=2 \pi / \omega$. The latter has a spatial dependence modelled by the factor $\textbf E_s(\textbf r)$. The electrostatic field causes a force $-q\nabla \phi$ on the particle, hence does not contribute to the ponderomotive force and is omitted here. The total magnetic field may be written as
\begin{equation}
	\textbf B = \textbf B_0 + \delta \textbf B  = B_0 \textbf e_z + \delta \textbf B \\
\end{equation}

\noindent We have introduced a local Cartesian coordinate system where the background magnetic field $\textbf B_0$  is aligned with the $z$-axis. With $\delta \textbf B$ we denote the magnetic field obtained from Faraday's law, due to spatial variations in $\textbf E_s$. 

\subsubsection*{Lower order particle motion}

The spatial electric field dependence is assumed small and will be neglected at lowest order. We obtain the corresponding low order equation of motion, with $\textbf r_0$ denoting  a \textit{fixed} spatial position,
\begin{equation}
	m \frac {d \textbf v} {dt} = q(\textbf E(\textbf r_0) + \textbf v \times \textbf B_0) .
\end{equation}

\noindent Separating into parallel and perpendicular (to $\textbf B_0$) motion so that $\textbf v = \textbf v_\parallel + \textbf v_\perp$ and writing $\textbf E(\textbf r_0) = \textbf E_s(\textbf r_0) \cos \omega t \equiv (\textbf E_\parallel + \textbf E_\perp)\cos \omega t$, we have
\begin{equation}
	\begin{split}
	m \frac {d \textbf v_\parallel} {dt} = q\textbf E_\parallel \cos \omega t  \\
	m \frac {d \textbf v_\perp} {dt} = q\textbf E_\perp \cos \omega t + q\textbf v_\perp \times \textbf B_0   \\
	\end{split}
\end{equation}

\noindent Denoting $\textbf v_\parallel = (0,0,v_z)$, $\textbf v_\perp = (v_x,v_y,0)$, and similarly for $\textbf E_\parallel$ and $\textbf E_\perp$, we obtain the solutions

\begin{equation}
	\begin{split}
	v_x = c_1 \cos \omega_c t + c_2 \sin \omega_c t - \frac{q}{m} \frac {\omega E_x \sin \omega t - \omega_c^* E_y \cos \omega t} {\omega_c^2-\omega^2} \\
	v_y = c_3 \cos \omega_c t + c_4 \sin \omega_c t - \frac{q}{m} \frac {\omega E_y \sin \omega t + \omega_c^* E_x \cos \omega t} {\omega_c^2-\omega^2} \\
	v_z = \frac {q}{m} \frac {E_z \sin \omega t} {\omega} + c_5  , \\
	\end{split}
\end{equation}

\noindent where we have introduced the cyclotron frequency $\omega_c = |q|B_0/m$. For simplifying formulas, we also introduced $\omega_c^* = qB_0/m$, being positive for ions and negative for electrons. Here $c_1-c_5$ are arbitrary constants of integration. The resonances for frequencies $\omega \approx \omega_c$ are particularly interesting. 

For transverse electromagnetic waves it holds that $\textbf E \perp \delta \textbf B$ and $\textbf E \perp \textbf k$, where $\textbf k$ is the wave vector or direction of propagation, so that we may assume the electric field vector to lie in a plane. We here set $(E_x,E_y,E_z) = (E_\perp, 0, E_\parallel)$. This holds for linearly polarized electric fields whereas, for circularly or elliptically polarized electric fields, our results will be approximative. Without loss of generality we may also set the phases of the velocities so that $c_2 = c_3 = c_5 = 0$. This results in

\begin{equation}
	\begin{split}
	v_x = c_1 \cos \omega_c t  - \frac{q}{m} \frac {\omega E_\perp \sin \omega t} {\omega_c^2-\omega^2}, \\
	v_y = c_4 \sin \omega_c t  -  \frac{q}{m} \frac {\omega_c^* E_\perp \cos \omega t} {\omega_c^2-\omega^2}, \\
	v_z = \frac {q}{m} \frac {E_\parallel \sin \omega t} {\omega}. \\
	\end{split}
\end{equation}

We have now solved Eq. (B4) for $\textbf v$, neglecting spatial variations of the oscillating electric field. The time behaviour of the components of $\textbf v$  is obviously oscillatory with frequencies $\omega$ and $\omega_c$. Physically, the particles rotate with frequency $\omega_c$ around the field lines of the background magnetic field $\textbf B_0$, perturbed by the perpendicular component of the oscillating electric field. The parallel electric field component causes parallel oscillations with frequency $\omega$. Time-averaging over several periods results in zero net velocity in the coordinate directions. To obtain a finite resultant motion we must go to higher order and include spatial variations in the electric field. 

\subsubsection*{Higher order particle motion}

To account for spatial variations, we need to perform a Taylor expansion of the electric field around the fixed spatial position $\textbf r_0$, about which the particle oscillates;
\begin{equation}
	\textbf E(\textbf r) = \textbf E(\textbf r_0) + (\delta \textbf r \cdot  \nabla)\textbf E_{\textbf r = \textbf r_0}
\end{equation}

\noindent where $\delta \textbf r = (\delta x,\delta y, \delta z) = \int \textbf v dt$. The force on the particle is obtained from the higher order equation of motion
\begin{equation}
	m \frac {d \textbf v_2} {dt} = q( (\delta \textbf r \cdot  \nabla)\textbf E + \textbf v \times \delta \textbf B) .
\end{equation}

\noindent This equation is second order in space, thus the notation $\textbf {v}_2$ for the spatial variable, since the terms on the right hand side depend on the particle position and velocity, respectively, as well as on the spatial variations of the electric and induced magnetic fields, respectively. 

We may now write, obtaining $\delta \textbf r$ from Eq. (B8);
\begin{equation}
	(\delta \textbf r \cdot  \nabla)\textbf E = \cos \omega t (\delta \textbf r \cdot  \nabla)\textbf E_s = \cos \omega t (\delta \textbf r \cdot  \nabla)(E_\perp,0,E_\parallel)
\end{equation}
\begin{equation}
	\begin{split}
	\delta \textbf r \cdot  \nabla = \Bigg(c_1 \frac {\sin \omega_c t} {\omega_c}  + \frac{q}{m} \frac { E_\perp \cos \omega t} {\omega_c^2-\omega^2}\Bigg)\frac {\partial}{\partial x} + \\
	\Bigg(- c_4 \frac{\cos \omega_c t} {\omega_c}  -  \frac{q}{m \omega} \frac {\omega_c^* E_\perp \sin \omega t} {\omega_c^2-\omega^2}\Bigg) \frac {\partial} {\partial y} +  \\
	\Bigg(- \frac {q}{m} \frac {E_\parallel \cos \omega t} {\omega^2}\Bigg) \frac {\partial} {\partial z}  \\
	\end{split}
\end{equation}

\noindent We are interested in the time-averaged force on the particle, that is

\begin{equation}
	< m \frac {d \textbf v_2} {dt}> = q( <(\delta \textbf r \cdot  \nabla)\textbf E > + <\textbf v \times \delta \textbf B>) .
\end{equation}

Defining the time-average of a quantity $Q$ as
\begin{equation}
	< Q > = \frac {1}{T} \int_0^T Q dt = \frac {\omega}{2\pi N} \int_0^{2\pi N/ \omega} Q dt
\end{equation}
\noindent the only non-zero terms of $< (\delta \textbf r \cdot  \nabla)\textbf E >$ in Eq. (B11), for $N \gg 1$ and $\omega \neq \omega_c$, are those that contain the factors $ < \cos^2 \omega t >  =  1/2$. Thus,
\begin{equation}
	< (\delta \textbf r \cdot  \nabla)\textbf E > =  \frac{q}{2m} \frac { E_\perp} {\omega_c^2-\omega^2} \frac {\partial}{\partial x}(E_\perp,0,E_\parallel)
	- \frac {q}{2m} \frac {E_\parallel} {\omega^2} \frac {\partial} {\partial z} (E_\perp,0,E_\parallel)
\end{equation}

For determining $<\textbf v \times \delta \textbf B>$ we note that 
\begin{equation}
	\delta \textbf B = -\frac {1}{\omega} \nabla \times \textbf E_s \sin \omega t
\end{equation}
obtains from Eqs. (B2)-(B4) by integrating in time. Furthermore
\begin{equation}
	\nabla \times \textbf E_s = \Bigg(\frac {\partial E_\parallel}{\partial y}, \frac {\partial E_\perp}{\partial z}-\frac {\partial E_\parallel}{\partial x}, -\frac {\partial E_\perp}{\partial y}\Bigg) .
\end{equation}

Using Eqs. (B8), (B16) and (B17), employing the fact that only terms with factors $<\sin^2 \omega t>=1/2$ give finite contributions, we can write
\begin{align}
	<\textbf v \times \delta \textbf B> = \frac{1}{\omega}\Bigg(&\frac {qE_\parallel} {2m\omega}\Bigg(\frac {\partial E_\perp}{\partial z}-\frac {\partial E_\parallel}{\partial x}\Bigg), \nonumber \\
	&\frac {-qE_\parallel}{2m\omega} \frac {\partial E_\parallel}{\partial y}+ 	\frac {q} {2m} \frac {\omega E_\perp}{\omega_c^2-\omega^2} \frac {\partial E_\perp}{\partial y}, \nonumber \\
	&\frac {q} {2m} \frac {\omega E_\perp}{\omega_c^2-\omega^2} \Bigg(\frac {\partial E_\perp}{\partial z} -  \frac {\partial E_\parallel}{\partial x}\Bigg)\Bigg)  
\end{align}

Gathering terms, we find
\begin{equation}
	<(\delta \textbf r \cdot  \nabla)\textbf E + \textbf v \times \delta \textbf B> = - \frac {q}{4m\omega ^2}\nabla E_\parallel^2 + \frac {q}{4m (\omega_c^2-\omega^2)} \nabla E_\perp^2
\end{equation}

Finally we obtain, from Eqs. (B13) and (B19),

\begin{equation}
	\textbf F_p =  - \frac {q^2}{4m\omega ^2}\nabla E_\parallel^2 - \frac {q^2}{4m (\omega^2-\omega_c^2)} \nabla E_\perp^2
\end{equation}

\noindent which is the ponderomotive force on a charged particle due to an oscillating electric field as modelled in Eq. (B3). This force is generally repelling for both electrons and ions in the sense that it points away from the region with higher electric field, which usually is the source electrode or antenna. The reason that the force acts in the same direction for electrons and ions is the following. In Eq. (B13) the electric force on the particle is proportional to $q$. It is also proportional to $\delta \textbf r$. From Eq. (B15) it is seen that the terms of $\delta \textbf r$ that contribute to the time-averaged force are also proportional to $q$. Thus, the first right hand term in Eq. (B13) is proportional to $q^2$. A similar argument holds for the second term of Eq. (B13), where the relevant terms of $\textbf v$ are proportional to $q$.   

It may, however, be noted that for strong $\nabla E_\perp^2$ in relation to $\nabla E_\parallel^2$, the ponderomotive force can become attractive when $\omega < \omega_c$. This is due to the change in velocity direction as seen in Eq. (B8).

If $\nabla E_\parallel^2 \ne \textbf 0$ the ponderomotive force increases in magnitude as $\omega^2 \rightarrow 0$. The reason for this is that for frequencies much lower than the cyclotron frequency, the particles have more time to respond to the oscillating field, leading to a larger displacement per oscillation. This results in a greater force pushing the particles away from regions of higher intensity. Mathematically, this is evidenced by Eqs. (B2)-(B4), (B8) and (B19). As the RF frequency $\omega$ decreases, the effect of inertia also decreases and the particle can follow the electric field to accelerate to a higher velocity $\textbf v$ along the magnetic field before turning back. Similarly, for a specified oscillating electric field, the magnitude of $\delta \textbf B$ becomes larger if the electric field oscillates at a slower pace $\omega$. Finally, the particle position $\delta \textbf r$ along the magnetic field develops with an $\omega^{-2}$ dependence, as seen from Eq. (B8). A combination of these effects yields an $\omega^{-2}$ dependence for the ponderomotive force. Conversely, at high frequencies, the particle cannot follow the rapid oscillations of the electric field effectively because of its inertia. As a result, the particle's motion averages out over a shorter period, and the ponderomotive force becomes weaker.
     
The derivation presented here shows that the physical reason for the ponderomotive force to appear only at second order in space is because, at this order, the particle motion and the applied electric and magnetic fields are in perfect phase, yielding a non-zero result when time averaged. At lower order the electric field in the equation of motion Eq. (B5) is computed at the fixed point $\textbf r_0$, thus not accounting for possible electric field gradients along the orbit of the particle, contributing to a resultant force.

The fact that $\omega_{ce} = eB/m_e$ is very large for electrons implies that the ponderomotive force for electrons, for $\omega \ll \omega_{ce}$, can be written
\begin{equation}
	\textbf F_{pe} =  - \frac {q^2}{4m_e\omega ^2}\nabla E_\parallel^2  .
\end{equation}

Due to the inverse mass dependence, the ponderomotive force on ions is generally much smaller than that on the electrons, but the resonance near the cyclotron frequency $\omega_{ci}$ can be utilized to produce a predominant ponderomotive force 
\begin{equation}
	\textbf F_{pi} =  - \frac {q^2}{4m_i (\omega^2-\omega_{ci}^2)} \nabla E_\perp^2  .
\end{equation}
This force is repulsive (with respect to the RF source) for $\omega > \omega_{ci}$.

\subsubsection*{Ponderomotive force near resonance $\omega = \omega_c$}
The ponderomotive force formula (B20) features a resonance at $\omega = \omega_c$. It is of interest to determine whether an expression for the ponderomotive force can be found for near-resonant frequencies. However, since interaction between the particle motion and the wave frequency leads to resonant energy absorption where the energy and Larmor radius of the particle increases rapidly additional physics, including collisional drag forces, are required for accurate modelling. This is shown as follows.

Setting $\omega = \omega_c$ in Eq. (B6) results in expressions for perpendicular particle motion that differ from those of Eq. (B8); this results in
\begin{equation}
	\begin{split}
	v_x = c_1 \cos \omega_c t  + \frac{q E_\perp}{2 m} t\cos \omega_c t,  \\
	v_y = c_4 \sin \omega_c t  - \frac{|q| E_\perp}{2 m} t\sin \omega_c t, \\
	v_z = \frac {q}{m} \frac {E_\parallel \sin \omega_c t} {\omega_c}. \\
	\end{split}
\end{equation}

The second terms in the expressions for $v_x$ and $v_y$ are oscillating, thus not contributing to average drifts. They grow, however, with time. This is because in the reference frame of the particle, at $\omega = \omega_c$ the applied electric RF field appears as an accelerating DC field. The rate of growth can be appreciated by assuming a DD plasma with temperature $T=20$ keV. For this case the amplitudes of $v_x$ and $v_y$ for ions would rise to thermal levels within a micro-second at electric field strengths of 58 kV/m. For electrons, the growth is nearly 2000 times faster. This shows that Eq. (B23) must be used with caution for RF frequencies near the cyclotron frequency of the particle species studied. It is furthermore clear that near-resonant tuning of the applied RF frequency may be costly due to unwanted heating of the particle species.


\end{document}